\documentclass{elsart}
\usepackage{epsfig} 

\newcommand{\beq}{\begin{equation}}
\newcommand{\eeq}{\end{equation}}
\newcommand{\beqa}{\begin{eqnarray}}
\newcommand{\eeqa}{\end{eqnarray}}
\newcommand{\om}{\Omega_m}
\newcommand{\calr}{{\mathcal R}}

\def\fun#1#2{\lower3.6pt\vbox{\baselineskip0pt\lineskip.9pt
  \ialign{$\mathsurround=0pt#1\hfil##\hfil$\crcr#2\crcr\sim\crcr}}}

\begin{document} 
\begin{frontmatter}
\title{Light Thoughts on Dark Energy} 
\author{Eric V.~Linder} 
\address{Physics Division, Lawrence Berkeley National 
Laboratory, Berkeley, CA 94720} 

\begin{abstract} 
The physical process leading to the acceleration of the expansion 
of the universe is unknown.  It may involve new high 
energy physics or extensions to gravitation.  Calling this  
generically dark energy, we examine the consistencies and relations 
between these two approaches, showing that an effective 
equation of state function $w(z)$ is broadly useful in describing 
the properties of the dark energy.  A variety of cosmological 
observations can provide important information on the dynamics of 
dark energy and the future looks bright for constraining dark energy, 
though both the measurements and the interpretation will be challenging. 
We also discuss a more direct relation between the spacetime geometry 
and acceleration, via ``geometric dark energy'' from the Ricci 
scalar, and superacceleration or phantom energy where the fate 
of the universe may be more gentle than the Big Rip. 
\end{abstract} 

\end{frontmatter}

\section{Introduction} \label{sec.intro}

The acceleration of the expansion of the universe poses a 
fundamental challenge to the standard models of both particle 
physics and cosmology.  There is no established framework for 
the new physics required, but we know that within Einstein's 
field theory we can treat modifications to either the 
right hand side -- energy-momentum components, e.g.\ a scalar 
field called quintessence -- or the left hand side -- the 
geometry of spacetime.  Certainly the cosmological constant 
$\Lambda$ is equally at home in either location. 

Here we examine what future observations can teach us about 
dark energy and the differences between a new physical component 
and an extension of gravitation theory.  We give a very brief 
summary of dark energy theory, moving on to dark energy 
phenomenology -- how to describe it in a manner amenable to both 
theory and observations, then dark energy fantasy -- can we 
constrain complexities in the description, and finally dark energy 
reality -- what we might expect to learn in the next decade. 

\section{Dark Energy Theory} \label{sec.theory}

Guidance on the nature of dark energy from fundamental theory has 
been loose.  While there is strong motivation for a cosmological 
constant, it encounters fine tuning and coincidence problems.  Attempts 
to overcome these have led to tracer and tracker fields.  The former, 
with a constant energy density relative to the dominant component, runs 
into problems with altered dynamics in the early universe, e.g.\ in 
the primordial nucleosynthesis era.  The latter, where the field 
follows an attractor trajectory despite starting from a wide variety 
of initial conditions, has difficulties reaching an equation of state 
pressure to energy density ratio $w\equiv p/\rho<-0.7$, while current 
observations favor $w<-0.78$ at 95\% confidence \cite{knop}. 

So theorists are currently in the state of trying anything and 
everything, including extra dimension models, tachyonic models, 
phase transition models, etc.  With such a plethora of interest and 
examination, perhaps tomorrow a compelling model will leap forward 
from astro-ph or hep-th and convince the community that this is 
the natural, well motivated theory to work with.  But it may be likelier 
that we will continue to have a surfeit of possibilities.  The solution 
to this is likely to be survival of the fittest, in the original 
Darwinian sense: those theories that fit the data best will continue. 
This leads to the point of phenomenology -- how to best interpolate 
between theory and observations so as to interpret the data robustly 
and cleanly. 

\section{Dark Energy Phenomenology} \label{sec.phenom}

Observations to date have basically been of the expansion dynamics of 
the universe, one or another proxy for the scale factor as a function 
of time, $a(t)$.  The Type Ia supernova method is the most direct 
tracer of this, with the redshift $z$ giving the scale factor and the 
luminosity giving the distance through the cosmological inverse square 
law, and hence the lookback time.  To link to the underlying theory, 
but in a more model independent way, we can employ the equation of 
state (EOS) variable $w(z)$ as intermediary between $a(t)$ and, say, the 
scalar field potential $V(\phi)$.  It is straightforward to carry 
out the translation between observations and a Lagrangian in theory, 
illustratively $V(\phi(a(t)))$. 

As mentioned, observations already constrain the recent, averaged EOS 
to lie within about 15\% (68\% cl) of the value $-1$.  But the physics 
of the dark energy lies in its dynamics, the variation $w(z)$.  Next 
generation experiments measuring distances and growth of structure 
should map $w(z)$ out to $z\approx1.7$, if systematic effects from 
observational uncertainties and other astrophysical variations can 
be tightly controlled.  One leading contender of a dark energy explorer 
is the Supernova/Acceleration Probe (SNAP: \cite{snap}), which is 
specifically designed for systematics control and employs both 
supernovae and weak gravitational lensing studies, as well as having 
further capabilities.  Such a mission may constrain a measure of the 
time variation $w'\equiv dw/d\ln a|_{z=1}$ to within 0.08.  The 
combination of high precision and high accuracy required certainly 
argues for a space mission. 

Since the dynamics of the expansion is the key element linking the 
observations to the theory, we can see from the Friedmann equations 
for the expansion, involving both $\dot a$ and $\ddot a$, that we 
need to take into account not only the dark energy density but also 
its pressure.  These can conveniently be combined in the equation of 
state ratio $w(z)$.  However, this parametrization has greater 
applicability as well.  A modification to the Friedmann equation 
can be translated to an effective parameter $w(z)$: 
\beq 
w(z)\equiv -1+\frac{1}{3}\frac{d\ln(\delta H^2/H_0^2)} 
{d\ln(1+z)}, 
\eeq 
where $\delta H^2$ is defined to be $H^2-(8\pi/3)\rho_m$, i.e.\ 
whatever is not matter in the equation for the expansion rate $H=\dot a/a$. 

This presents a wonderfully inclusive and model independent way to 
treat a large variety of dark energy theories, within the same 
``language''.  Of course, perhaps the resulting function $w(z)$ may 
look complicated, but this does not affect the mathematical treatment. 
And in fact, one finds that a remarkably simple but robust parametrization 
\beq 
w(z)=w_0+w_a(1-a)=w_0+w_az/(1+z), 
\eeq 
originally developed for slow roll scalar fields, works extremely well 
for many extensions to gravitation and high energy physics.  

Figure 1 (left panel) illustrates this for an extra dimension braneworld 
model and 
a vacuum metamorphosis phase transition model.  Note that these two 
theories span extremes in the sense of having averaged EOS near the upper 
and lower limits of current data, and possessing opposite signs of the 
time variation.  Observations of both the growth history of structure 
in linear perturbation theory, written in terms of the gravitational 
potential $\Phi(z)$, and the expansion history in terms of the supernova 
magnitude-redshift relation (not shown, but fit to 0.01 mag out to $z=2$), 
can be superbly fit by $(w_0,w_a)$ with values of (-0.78,0.32) and 
(-1,-3) respectively.  This simplicity is a feature, not a bug, of the 
observations involving integrals over the EOS; that is, one does {\it not\/}  
attempt to reconstruct $w(z)$ (in particular in the case of vacuum 
metamorphosis this is not a good fit), but rather to match the observations 
going forward from the parametrization. 

\begin{figure}[!hbt]
\begin{center} 
\psfig{file=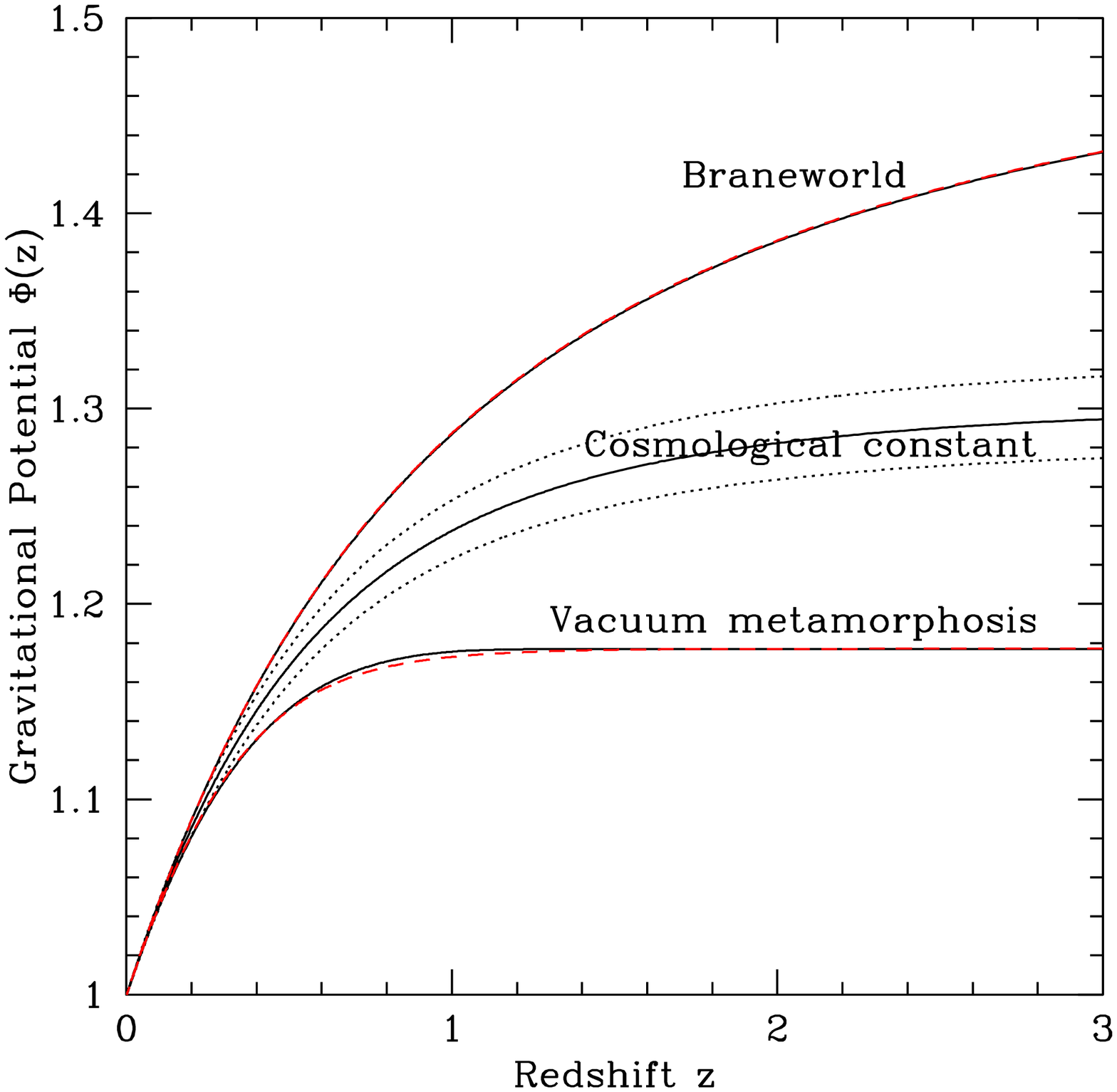,width=6.8cm} 
\psfig{file=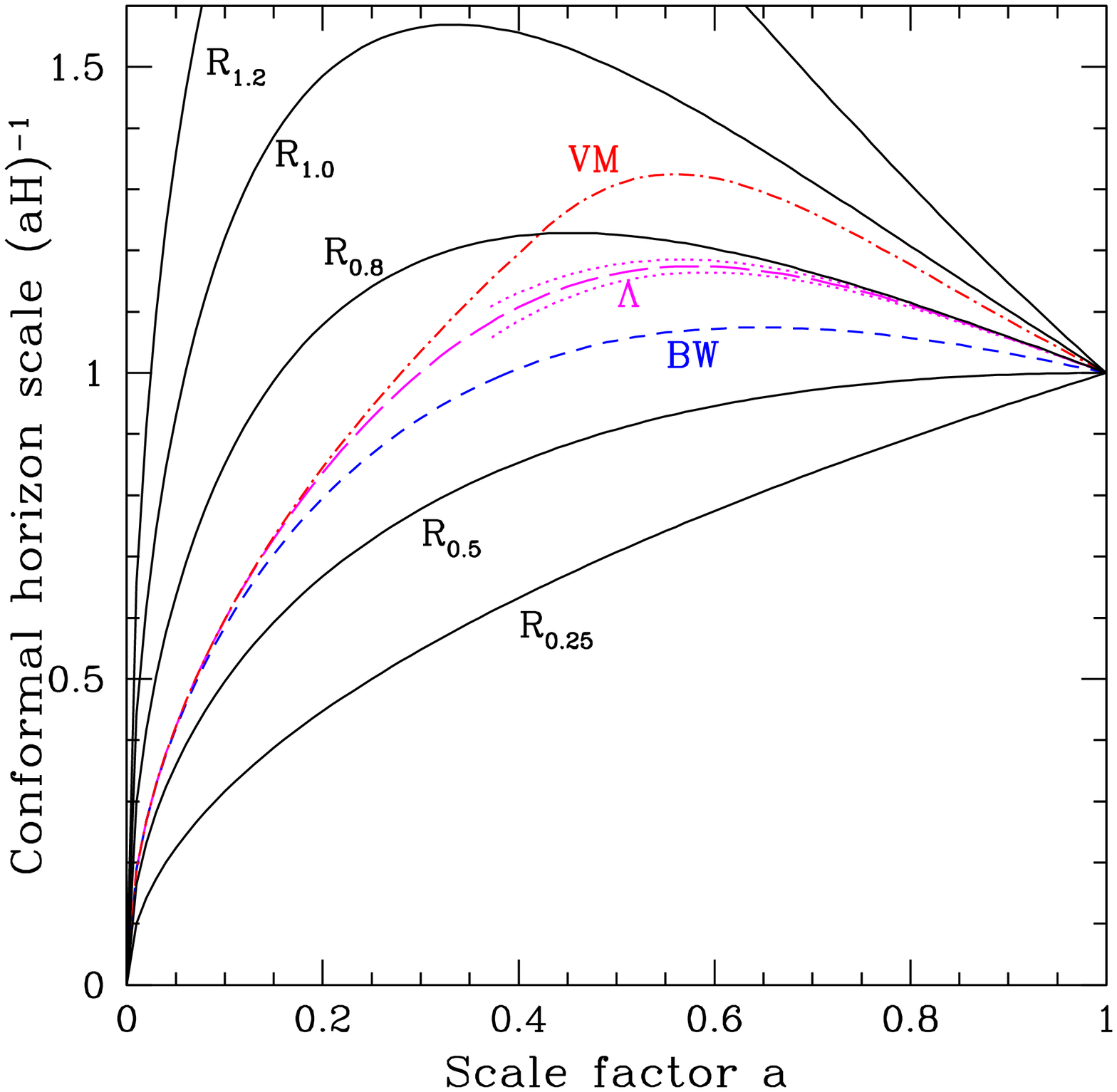,width=6.8cm} 
\caption{[Left panel] The gravitational potential $\Phi(z)$ for two 
non-quintessence models shows the decay of the potential as the 
expansion accelerates.  Dashed, red curves are for the mimicking 
quintessence models.  Dotted outliers to the cosmological constant 
curve show the effect of changing the matter density $\om$ by 0.02. 
[Right panel] The expansion history is plotted in terms of conformal horizon 
scale vs.\ scale factor for various modified gravity and spacetime geometry 
models.  The Ricci geometric dark energy models (solid, black curves) are 
subscripted with the present value of $\calr$.  Negative slopes 
indicate an accelerating epoch while slopes more steeply negative than 
a critical value ($-1$ at the present) indicate superacceleration.  
} 
\end{center} 
\end{figure}

Another possibility besides modifying the Friedmann equation is to 
attempt to address the acceleration directly through its relation to 
spacetime curvature, resident in the Principle of Equivalence.  We 
can define a ``geometric dark energy'' based on the Ricci scalar 
curvature $R$, with $\calr=R/(12H^2)$ a key parameter in describing 
acceleration.  This is related to parametrizations in terms of the 
second and third derivatives of the scale factor (e.g.\ ``statefinder'' 
from \cite{starosahni} and ``jerk'' from \cite{visser}) but gives 
a coherent foundation in terms of a geometric quantity rather than a 
Taylor expansion. 

When $\calr>1/2$ then the expansion of the universe accelerates and 
when $\calr>1$ it superaccelerates.  To make a close analogy with 
inflation theory -- acceleration in the early universe -- consider the 
conformal diagram in the right panel of Figure 1.  A negative slope to the 
conformal horizon scale $(aH)^{-1}$ with respect to $a$ means that comoving 
scales leave the horizon: just like in inflation a mark of acceleration. 
We see that this is achieved at the present epoch ($a=1$) by the 
curve marked $R_{0.5}$, with $\calr=1/2$ today.  Superacceleration 
occurs for those models with curves steeper than $R_{1.0}$.  We 
conjecture that enhanced particle production from the Rindler horizon 
created by superacceleration can obviate the Big Rip and create a 
cyclic scenario \cite{grav}. 

\section{Dark Energy Fantasy} \label{sec.fantasy}

If phenomenology describes the properties we expect the dark energy 
to possess, yet we must be aware of the possibility of more complicated 
situations where the basic characteristics of dark energy darkness, 
smoothness, and determination of the fate of the universe may not be 
so simple.  Darkness can be lifted by self interactions or couplings to 
matter (which face stringent tests from fifth force laboratory 
experiments and astrophysical measurements of the matter power spectrum 
shape, e.g.\ \cite{sandvik}).  Couplings to gravitation, e.g.\ in 
scalar-tensor theories of 
gravity, allow dark energy clumping on subhorizon scales, the possibility 
of backreaction from nonlinear structure formation, and maybe even a 
solution to the fine tuning problem through an attractor trajectory 
\cite{rboost}. 

Without a complete theory of dark energy we cannot be sure of claims 
that measuring its density and time variation over a limited range of 
redshifts teaches us enough about dark energy to predict the fate of 
the universe.  Even simple dark energy models, such as the linear 
potential, can have a currently accelerating phase that then gives 
way to a deceleration and collapse in a finite time \cite{linde}. 

\section{Dark Energy Reality} \label{sec.reality}

The first step to learning about the new fundamental physics behind 
dark energy is measuring the time variation $w'=w_a/2$, 
with next generation experiments like SNAP.  Before then, we may narrow 
in on a time averaged quantity $\langle w\rangle$, but this will 
mostly generate papers and speculation, not motivated theories.  We 
have seen that even $w_0$ and $w_a$ go a long way toward fitting 
observations in terms of a real or effective equation of state function 
$w(z)$, and that $w(z)$ is an extremely general language for talking 
about the underlying physics. 

Given this language, survival of the fittest enters, with the data 
deciding how or whether to go beyond a cosmological constant $\Lambda$. 
This will require careful and challenging levels of precision and 
systematics control, and needs to be complemented and crosschecked by 
multiple cosmological methods. 

Part of the reality might be what lies in the future beyond $w(z)$. 
Consider the analogy with inflation.  Inflation is not simply deSitter 
expansion: we {\it want\/} complexity -- perturbations, running, tensor 
modes -- to learn deeper physics.  Similarly for dark energy we learn 
much through $w'$ but there may detectable (some day) noncanonical 
sound speed $c_s^2$, inhomogeneities $\delta\phi$, couplings\dots\ 
Exciting times lie ahead for dark energy research.

\begin{ack}         
This work has been supported 
in part by the Director, Office of Science, Department of Energy under 
grant DE-AC03-76SF00098.  I thank David Cline for inviting me to present 
this overview of dark energy theory at the UCLA Dark Matter/Dark Energy 2004 
meeting, and Robert Caldwell, Varun Sahni, Alexei Starobinsky, and Yun 
Wang for inspiring greater clarity of thought on various aspects. 
\end{ack}

\end{document}